\documentclass[letterpaper, 10 pt, conference]{ieeeconf}  

\IEEEoverridecommandlockouts                             
\usepackage{balance}
\usepackage{silence}
\IEEEoverridecommandlockouts

\usepackage{cite}
\usepackage{graphicx}
\usepackage{picinpar}
\usepackage{stfloats}
\usepackage{url}
\usepackage{flushend}
\usepackage[latin1]{inputenc}
\usepackage{colortbl}
\usepackage{soul}
\usepackage{multirow}
\usepackage{pifont}
\usepackage{alltt}
\usepackage{enumerate}
\usepackage{siunitx}
\usepackage{pbox}
\usepackage{amsmath,amssymb,amsfonts}
\usepackage{amsthm}
\usepackage{algorithmic}
\usepackage{algorithm}
\usepackage{bm}
\usepackage{graphbox}
\usepackage{textcomp}
\usepackage{xcolor}
\usepackage[export]{adjustbox}
\usepackage{color}
\usepackage{subcaption}
\usepackage{caption}
\usepackage{tikz}
\def\BibTeX{{\rm B\kern-.05em{\sc i\kern-.025em b}\kern-.08em
    T\kern-.1667em\lower.7ex\hbox{E}\kern-.125emX}}

\definecolor{LightCyan}{rgb}{0.8,0.8,1.0}
\definecolor{LightRed}{rgb}{1.0,0.8,0.8}
\definecolor{LightGreen}{rgb}{0.8,1.0,0.8}
\definecolor{LightYellow}{rgb}{1.0,1.0,0.8}

\newcommand{\blue}[1]{\color{black}{#1}\color{black}\phantom{}}

\newtheorem{definition}{Definition}

\newtheorem{problem}{Problem}

\ActivateWarningFilters[balance]

\makeatletter
\let\NAT@parse\undefined
\makeatother
\usepackage[hidelinks]{hyperref}

\newcommand{\bfW}{\mathbf{W}}

\title{\LARGE \bf
Learning to Identify Graphs from Node Trajectories \\ in Multi-Robot Networks}

\author{Eduardo Sebasti\'{a}n \and Thai Duong \and Nikolay Atanasov \and Eduardo Montijano \and Carlos Sag\"{u}\'{e}s%
\thanks{E. Sebasti\'{a}n, E. Montijano and C. Sag\"{u}\'{e}s are with the RoPeRt group, at DIIS - I3A, Universidad de Zaragoza, Spain (e-mails: \texttt{\small \{esebastian, emonti, csagues\}@unizar.es}).}%
\thanks{T. Duong and N. Atanasov are with the Department of Electrical and Computer Engineering, University of California San Diego, La Jolla, CA 92093 USA (e-mails: \texttt{\small \{tduong, natanasov\}@ucsd.edu}).}%
\thanks{This work has been supported by NSF CCF-2112665 (TILOS) and via Spanish projects PID2021-125514NB-I00, PID2021-124137OBI00 and TED2021-130224B-I00 funded by MCIN/AEI/10.13039/501100011033, by ERDF A way of making Europe and by the European Union NextGenerationEU/PRTR, DGA T45-23R, and Spanish grant FPU19-05700 and EST22/00253.}%
}

\newcommand\copyrighttext{%
  \footnotesize \textcopyright This paper has been accepted for publication in the 4th IEEE International Symposium on Multi-Robot and Multi-Agent Systems (IEEE MRS 2023). Please cite the paper as: E. Sebasti\'{a}n, T. Duong, N. Atanasov, E. Montijano and C. Sag\"{u}\'{e}s,``Learning to Identify Graphs from Node Trajectories in Multi-Robot Networks'', 4th IEEE International Symposium on Multi-Robot and Multi-Agent Systems (MRS), 2023.}
\newcommand\copyrightnotice{%
\begin{tikzpicture}[remember picture,overlay]
\node[anchor=south,yshift=10pt] at (current page.south) {\fbox{\parbox{\dimexpr\textwidth-\fboxsep-\fboxrule\relax}{\copyrighttext}}};
\end{tikzpicture}%
}

\begin{document}

\maketitle
\copyrightnotice
\thispagestyle{empty}
\pagestyle{empty}

\begin{abstract}
The graph identification problem consists of discovering the interactions among nodes in a network given their state/feature trajectories. This problem is challenging because the behavior of a node is coupled to all the other nodes by the unknown interaction model. Besides, high-dimensional and nonlinear state trajectories make \blue{it} difficult to identify if two nodes are connected. Current solutions rely on prior knowledge of the graph topology and the dynamic behavior of the nodes, and hence, have poor generalization to other network configurations. To address these issues, we propose a novel learning-based approach that combines (i) a strongly convex program that efficiently uncovers graph topologies with global convergence guarantees and (ii) a self-attention encoder that learns to embed the original state trajectories into a feature space and predicts appropriate regularizers for the optimization program. In contrast to other works, our approach can identify the graph topology of unseen networks with new configurations in terms of number of nodes, connectivity or state trajectories. We demonstrate the effectiveness of our approach in identifying graphs in multi-robot formation and flocking tasks. 
\end{abstract}


\section{Introduction}\label{sec:intro}


The study of networked systems is essential in many
disciplines like brain network imaging \cite{Monti_NeuorImage_2014_NeuroscienceGraphLearning,Huang_ProceedingsIEEE_2018_SurveyGSPFunctionalBrainImaging, nozari2022structural}, genetics \cite{Julius_IETBS_2009_BiologyExample,Nabi-Abdolyousefi_TAC_2012_NodeKnockoutGraphLearning}, power networks \cite{Giannakis_ProceedingsIEEE_2018_SurveyTopologyIdentificationNonlinearitiesDynamics}, social networks \cite{Dong_TSP_2016_LearningSmoothLaplacians,Giannakis_ProceedingsIEEE_2018_SurveyTopologyIdentificationNonlinearitiesDynamics,Shafipour_ICASSP_2017_TopologyInferenceNonStationaryGraph,montijano2020distributed}, environmental monitoring \cite{ Pasdeloup_TSIPN_2017_CharacterizationGraphDiffusionStationary,Dong_TSP_2016_LearningSmoothLaplacians,Thanou_TSIPN_2017_LearningHeatDiffusionGraphs}, or general large-scale physically interconnected systems \cite{Timme_PRL_2007_RevealingConnectivityLinearSystems,Napoletani_PRE_2008_SparseGraphIdentification}. 
The interactions among entities play a central role in understanding networked systems and motivates an extensive effort to identify the interactions (i.e., the graph topology) from data \cite{Dong_SPM_2019_SurveyLearningGraphsFromData, mateos2019connecting, Xia_TAI_2021_GraphLearningSurvey}. 
For example, graph topology identification is crucial for modeling multi-robot interactions \cite{rossi2021multi} in learning collaborative behaviors from observations or demonstrations \cite{shi2020neural,jiahao2022learning,sebastian2023lemurs}. In this paper, we aim to develop an algorithm to identify the underlying graph topology that best describes the behavior of a networked system given its node trajectories.

A widely used approach for graph topology identification is graph signal processing~\cite{Ortega_IEEEProceedings_2018_GraphSignalProcessingSurvey}. Diffusion-based methods~\cite{Thanou_TSIPN_2017_LearningHeatDiffusionGraphs, Giannakis_ProceedingsIEEE_2018_SurveyTopologyIdentificationNonlinearitiesDynamics} assume that the node signals diffuse through the edges following heat kernels. The proposed approaches are posed as minimization problems over the heat kernels' parameters and the Laplacian matrix. However, the constraints on Laplacian matrix are non-convex, and an a priori dictionary of functions is required to find the kernels' parameters.
Another line of research focuses on the reconstruction of a graph shift operator by means of its eigenvalue and eigenvector pairs, under perfect observations and known eigenvectors \cite{Pasdeloup_TSIPN_2017_CharacterizationGraphDiffusionStationary,Segarra_TSIPN_2017_NetworkTopologyInferenceSpectralTemplates} or linear time-invariant single-input single-output nodes \cite{Shahrampour_TAC_2014_TopologyIdentificationDirectedDynamicalNetworks}.
A common issue is how to encode the constraints of adjacency and Laplacian matrices \cite{Segarra_TSP_2016_BlindIdentificationGraph}, which lead to NP-hard problems. Learning a Laplacian can be recast to a L1 norm minimization by exploiting a smoothness assumption \cite{Kalofolias_AIS_2016_LearnGraphSmoothSignals,Kalofolias_ICASSP_2017_LearningTimeVaryingGraphs}. This perspective has been extended to develop fast online algorithms \cite{Saboksayr_EUSIPCO_2021_OnlineGraphLearningSmoothnessPriors,Saboksayr_SPL_2021_AcceleratedGraphLearning} which benefit from the fact that the L1-norm and the Laplacian constraints can be rewritten as a vectorized multiplication and an indicator function, leading to unconstrained convex problems \cite{beck2014fast}. Nevertheless, these algorithms assume scalar node signals, which is not the usual case in general networked systems. 

A promising alternative to graph signal processing is machine learning algorithms that use attention~\cite{Vaswani_NIPS_2017_Attention} and self-attention~\cite{Shaw_arXiv_2018_SelfAttentionRelative} mechanisms. Self-attention discovers the relationships among elements of a sequence by computing an attention map, and its structure can be combined with linear layers and activation functions to encode nonlinear behaviors. Furthermore, attention layers allow for a time-varying size of one of the input dimensions, e.g., the number of robots in a robotic team. Regarding multi-robot systems, self-attention layers can be found in recent path planning~\cite{Kamra_NIPS_2020_MultiAgentSelfAttention,Li_RAL_2021_MessageAwareSelfAttention} or task scheduling~\cite{Wang_RAL_2020_SelfAttentionScheduling} applications. It is worth mentioning that there are learning techniques related to graphs which cannot be applied to our problem because they assume a known graph: relation prediction, graph regression,  clustering~\cite{Hamilton_Book_2020_GraphLearning}, and graph neural network solutions~\cite{Wang_ACMTG_2019_DynamicGCNNPointCloud,Goyal_KBS_2020_Dyngraph2vec}. 

We formulate the problem of identifying the graph topology of a networked system from state trajectories of its nodes (Section~\ref{sec:problem_statement}). Our main contribution is a graph topology identification approach (Section~\ref{sec:modules}) that captures high-dimensional node features in a strongly convex optimization problem for weighted adjacency matrix optimization. In contrast with learning-based techniques, our approach exploits principled strongly convex optimization to generate weighted adjacency matrices with guarantees of convergence. Compared to graph signal processing techniques, our approach relies on a self-attention-based neural networks to represent high-dimensional node features for the adjacency matrix optimization problem. The neural networks are trained to balance between the loss function and the regularizers in the objective function so that it best captures the sparsity of the graph. The approach is validated through multi-robot formation and flocking experiments in Section~\ref{sec:results}, and the the benefits of our proposal are discussed in Section~\ref{sec:conclusions}.


\section{Problem Formulation}\label{sec:problem_statement}
Consider a networked system characterized by an undirected weighted graph $\mathcal{G} = (\mathcal{V},\mathcal{E},\mathbf{W})$. The set of nodes is $\mathcal{V} = \{1,\hdots ,n\}$, with $n > 1$ the number of nodes. The interactions among nodes are represented by the set of edges $\mathcal{E} \subseteq \mathcal{V} \times \mathcal{V}$. The weighted adjacency matrix $\mathbf{W} \in \mathbb{R}^{n \times n}$ represents the intensity of the interactions among nodes, and it is such that $[\mathbf{W}]_{ij} = w_{ij} \in \mathbb{R}_{>0}$ if $(i,j) \in \mathcal{E}$ and $[\mathbf{W}]_{ij} = 0$ otherwise. We assume that there are no self-loops, i.e., $w_{ii}=0$. The set of neighbors of node $i$ is $\mathcal{N}_i = \{j\in \mathcal{V} | (i,j) \in \mathcal{E}\}$. Since $\mathcal{G}$ is undirected, $(i,j) \in \mathcal{E}$ implies that $(j,i) \in \mathcal{E}$ and $w_{ij} = w_{ji}$. We define the edge density as $\rho(\mathcal{G})=|\mathcal{E}|/n^2$, with $|\mathcal{E}|$ the number of edges. The (weighted) Laplacian is $\mathbf{L} = \mathsf{diag}(\mathbf{W} \mathbf{1}_n) - \mathbf{W}$, where $\mathbf{1}_n$ is the column vector of ones of size $n$. 

Node $i$ is characterized by a state $\mathbf{x}_i(t) \in \mathcal{X}_i \subseteq \mathbb{R}^{s}$ at a discrete time $t \in \mathbb{N}$, where $\mathcal{X}_i$ is the space of admissible states of dimension $s \in \mathbb{N}$. Each node obeys unknown discrete-time dynamics,
\begin{equation}\label{eq:dynamics}
    \mathbf{x}_i(t+1) = f_i(\mathbf{x}_i(t), \mathbf{x}_{\mathcal{N}_i}(t)),
\end{equation}
where $\mathbf{x}_{\mathcal{N}_i}(t) = \{\mathbf{x}_j(t)\}_{j\in\mathcal{N}_i}$ is the state of the neighbors of node $i$ at time $t$. Let $\mathbf{x}_i^{d}(t) = [\mathbf{x}_i(t-d), \hdots, \mathbf{x}_i(t)]$ be the trajectory formed by the last $d$ states of node $i$ at instant $t$. We define the tensor $\mathbf{X}(t) = [\mathbf{x}_1^{d}(t), \hdots, \mathbf{x}_{n}^{d}(t)]$ as the $n \times s \times d$ collection of trajectories at instant $t$. We assume that these trajectories are available at each instant, and the associated $\mathbf{W}$ is time-invariant. The aim of this work is to develop an approach to identify $\mathbf{W}$ given $\mathbf{X}(t)$.

We assume that the evolution of the node states is such that the graph state is smooth. Roughly speaking, the smoothness assumption implies that the neighbors of node $i$ have more similar values of $\mathbf{x}_i(t)$ compared to non-neighboring nodes. Formally, the total variation of the graph state is
\begin{equation}\label{eq:Omega}
    \varphi(\mathbf{x}(t)) = \mathbf{x}^{\top}(t)(\mathbf{L} \otimes \mathbf{I}_s)\mathbf{x}(t),
\end{equation}
where $\mathbf{x}(t) = [\mathbf{x}_1^{\top}(t),\hdots,\mathbf{x}_{n}^{\top}(t)]^{\top}$, $\otimes$ is the Kronecker product, and $\mathbf{I}_s$ is the identity matrix of dimension $s$. Then, we define graph state smoothness as follows.

\begin{definition}\label{def:smooth}
The state trajectories of graph $\mathcal{G}$ are smooth if $\varphi(\mathbf{x}(t))$ satisfies $\varphi(\mathbf{x}(t)) < \sigma$, with $0 < \sigma \ll \varrho$ and $\varrho \in \mathbb{R}_{>0}$ an upper bound on the total variation of the graph state. 
\end{definition}

We assume that the evolution of the node trajectories given by Eq.~\eqref{eq:dynamics} is such that graph state smoothness holds, which is the case for many multi-agent and multi-robot problems such as flocking, cooperative exploration, opinion dynamics, and consensus \cite{olfati2006flocking,atanasov2015decentralized,acemoglu2011opinion,kia2019tutorial}. The particular values for $\sigma$ and $\varrho$ depend on the application. Under these assumptions,
the goal of the paper is formulated as follows.


\begin{problem}\label{problem:uncompressed}
Given smooth node state trajectories $\mathbf{X}(t)$ obtained from a graph $\mathcal{G}$ with unknown dynamics $f_i(\bullet)$ $\forall i$ and edges $\mathcal{E}$, find the weighted adjacency matrix $\mathbf{W}$ of $\mathcal{G}$.
\end{problem}

\section{Learning to Identify Graphs}\label{sec:modules}


In this section, we propose an approach to solve Problem~\ref{problem:uncompressed} that combines (i) a fast strongly convex optimization algorithm with global convergence guarantees and (ii) a self-attention encoder that transforms high-dimensional node states to one-dimensional features and learns the regularization parameters that match the graph sparsity pattern. We describe the optimization algorithm in Section \ref{subsec:accelerated} and develop the self-attention encoder in Section \ref{subsec:SAA}. In Section \ref{subsec:proposal}, we describe our overall approach for graph topology learning from node trajectories.

\subsection{Accelerated graph learning from smooth signals}\label{subsec:accelerated}

In a networked system, the behavior of one node is influenced by the entire network by means of local interactions with neighboring nodes via Eq.~\eqref{eq:dynamics}. Hence, to identify the existence of interactions among two nodes
we need to consider, at least initially, the state information from all the nodes. In this section, we describe an optimization method \cite{Saboksayr_SPL_2021_AcceleratedGraphLearning} for scalar states ($s=1$) that finds the underlying graph that \blue{best describes the state trajectories} of the nodes. When state $\mathbf{x}_i$ is scalar, the trajectory $\mathbf{X}(t)$ is an $n \times d$ matrix. 



First, we define $\mathbf{Y}(t) \in \mathbb{R}^{n \times n}$ as the Euclidean distance matrix among the node states such that \mbox{$[\mathbf{Y}(t)]_{ij} = ||\mathbf{x}_i^d(t) - \mathbf{x}_j^d(t))||_2^2$ $\forall i,j \in \mathcal{V}(t)$}. Then, 
\begin{equation}\label{eq:equivalent_smoothness}
    \kern -0.2cm\varphi(\mathbf{x}(t)) = \hbox{trace}(\mathbf{X}^{\top}(t)\mathbf{L}\mathbf{X}(t)) = \frac{1}{2}||\mathbf{W} \circ \mathbf{Y}(t)||_1,
\end{equation}
where $\circ$ denotes element-wise product and $\|\cdot\|_1$ is the L1 norm. We omit the time dependence going forward to simplify the notation because all operations refer to the same instant $t$. Based on Eq.~\eqref{eq:equivalent_smoothness}, finding $\mathbf{W}$ translates into solving a convex inverse problem~\cite{Kalofolias_AIS_2016_LearnGraphSmoothSignals}:
\begin{subequations}\label{eq:opt1}
\begin{alignat}{2}
\underset{\mathbf{W}}{\min} & \hbox{ } ||\mathbf{W} \circ \mathbf{Y}||_1  -  \alpha \mathbf{1}_n^{\top} \log(\mathbf{W}\mathbf{1}_n) + \beta ||\mathbf{W}||_F^2   \label{eq:opt1cost}
\\
\text{s.t.} \hbox{ } &\; w_{ii} = 0 \hbox{, } w_{ij} \geq 0, \;\; \forall i,j \in \mathcal{V},\label{eq:opt1constraint1}
\end{alignat}
\end{subequations}
where $\alpha, \beta > 0$ are tuning parameters such that $\alpha$ penalizes the possibility of isolated nodes and $\beta$ encourages graph sparsity. To solve \eqref{eq:opt1} efficiently and with global convergence guarantees, Saboksayr and Mateos \cite{Saboksayr_SPL_2021_AcceleratedGraphLearning} proposed a dual reformulation of \eqref{eq:opt1}. Since the graph is undirected, we define $\mathbf{w}= \mathsf{vech}(\mathbf{W})$ as the half-vectorization of $\mathbf{W}$. Let \mbox{$\mathbf{y}=\mathsf{vec}(\mathbf{Y})$} be the vectorization of $\mathbf{Y}$ and $\mathbb{I}$ be the indicator function such that $\mathbb{I}\{\mathbf{w} \geq 0\} = 0$ and $\infty$ otherwise, where the operation $\geq$ is applied element-wise. Finally, $\mathbf{S} \in \{0,1\}^{n \times \frac{n(n-1)}{2}}$ is the operator that satisfies $\mathbf{W}\mathbf{1}_n = \mathbf{S}\mathbf{w}$. This leads to the unconstrained optimization problem
\begin{equation}\label{eq:opt2}
 \underset{\mathbf{w}}{\min} \kern 0.3cm \mathbb{I}\{\mathbf{w} \geq 0\} + 2\mathbf{w}^{\top}\mathbf{y} + \beta||\mathbf{w}||_2^2  - \alpha \mathbf{1}^{\top}_n \log(\mathbf{S}\mathbf{w}) .  
\end{equation}

Eq.~\eqref{eq:opt2} can be solved by accelerated dual proximal gradient methods with guarantees of global convergence, \blue{proved by Beck and Teboulle \cite{beck2014fast} in the general case, and by Saboksayr and Mateos \cite{Saboksayr_SPL_2021_AcceleratedGraphLearning} in the specific case of Eq.~\eqref{eq:opt2}}. In particular, we exploit the analytical expressions derived in \cite{Saboksayr_SPL_2021_AcceleratedGraphLearning}, summarized in Algorithm~\ref{al:smooth}. 

\begin{algorithm}
\caption{Accelerated dual proximal gradient method for graph identification}\label{al:smooth}
\begin{algorithmic}[1]
\STATE \textbf{Inputs}: $\mathbf{y}$, $\mathbf{S}$
\STATE \textbf{Initialization}: $\bm{\omega}_0 = \bm{\lambda}_0 = \bm{\lambda}_{-1} \sim \mathcal{U}(\frac{n(n-1)}{2})$, $\tau_0 = 1$
\STATE \textbf{Parameters}: $K  > 0$, $\alpha > 0$, $ \beta > 0$, $L = \frac{n-1}{\beta}$, $\epsilon > 0$  
\FOR{$k \in \{0, \hdots, K-1\}$}
    \STATE $\mathbf{w}_k = \max \left(\mathbf{0}, \displaystyle\frac{\mathbf{S}^{\top}\bm{\omega}_k - 2\mathbf{y}}{2\beta} \right)$
    \STATE $\mathbf{u}_k = \frac{1}{2}\left(\mathbf{S}\mathbf{w}_k - L \bm{\omega}_k + \sqrt{(\mathbf{S}\mathbf{w}_k - L \bm{\omega}_k)^2 + 4 \alpha L \mathbf{1}_n} \right)$
    \STATE $\bm{\lambda}_k = \bm{\omega}_k - L^{-1}(\mathbf{S}\mathbf{w}_k - \mathbf{u}_k)$
    \STATE $\tau_{k+1} = \frac{1 + \sqrt{1+4\tau_{k}^2}}{2}$
    \STATE $\bm{\omega}_k = \bm{\lambda}_k + \frac{\tau_{k} - 1}{\tau_{k+1}}(\bm{\lambda}_k - \bm{\lambda}_{k-1})$
    \STATE \textbf{if }$||\bm{\lambda}_k - \bm{\lambda}_{k-1}||/||\bm{\lambda}_{k-1}||<\epsilon$ \textbf{then break}
\ENDFOR 
\STATE \textbf{Output}: $\mathbf{w}_k$ 
\end{algorithmic}
\end{algorithm}

The algorithm iterates until convergence or a certain number of iterations $K \in \mathbb{N}$ is reached. The current iteration is denoted by $k$ and $\bm{\lambda}_k$ denotes the Lagrange multipliers of the dual optimization problem of Eq. \eqref{eq:opt2}. Algorithm~\ref{al:smooth} alternates between weighted adjacency matrix and Lagrange multipliers' updates. Once the algorithm terminates, the graph adjacency matrix is recovered from $\mathbf{w}_k$. \blue{Interestingly, the stability of the training process is not affected by the $\max$ operator in line 5 of Algorithm~\ref{al:smooth}. At the boundary $\frac{\mathbf{S}^{\top}\bm{\omega}_k - 2\mathbf{y}}{2\beta} = \mathbf{0}$, the gradient can be smooth by, e.g., using an approximated value like in Pytorch \cite{paszke2019pytorch}.}

Besides the global convergence\blue{\cite{Saboksayr_SPL_2021_AcceleratedGraphLearning}, which implies robustness against initializations}, Algorithm \ref{al:smooth} is efficient to compute so, at each instant, sufficient iterations can be run to ensure convergence to the weighted adjacency matrix that best describes the node state trajectories defined over the Euclidean space. 

\subsection{Self-attention encoder}\label{subsec:SAA}

To use Algorithm~\ref{al:smooth} for graph identification, the node state trajectories must be encoded to a feature space of dimension $s=1$. On the other hand, the cost function in Eq.~\eqref{eq:opt2} is taken with respect to the distance between state trajectories in Euclidean space, so the solution might not reflect the best topology given $\mathbf{X}(t)$ because the interactions may be determined by state proximity in a different space. Therefore, the encoding must be such that the Euclidean distance is the one that best reflects the distances among node trajectories. In addition, the encoding must handle a time-varying number of nodes and adjust to changes in the graph connectivity. To address all these points, in this section we develop a neural network encoder using self-attention \cite{Vaswani_NIPS_2017_Attention} to extract suitable features from the node state trajectories.

\begin{figure}
    \centering
    \includegraphics[width=0.9\columnwidth]{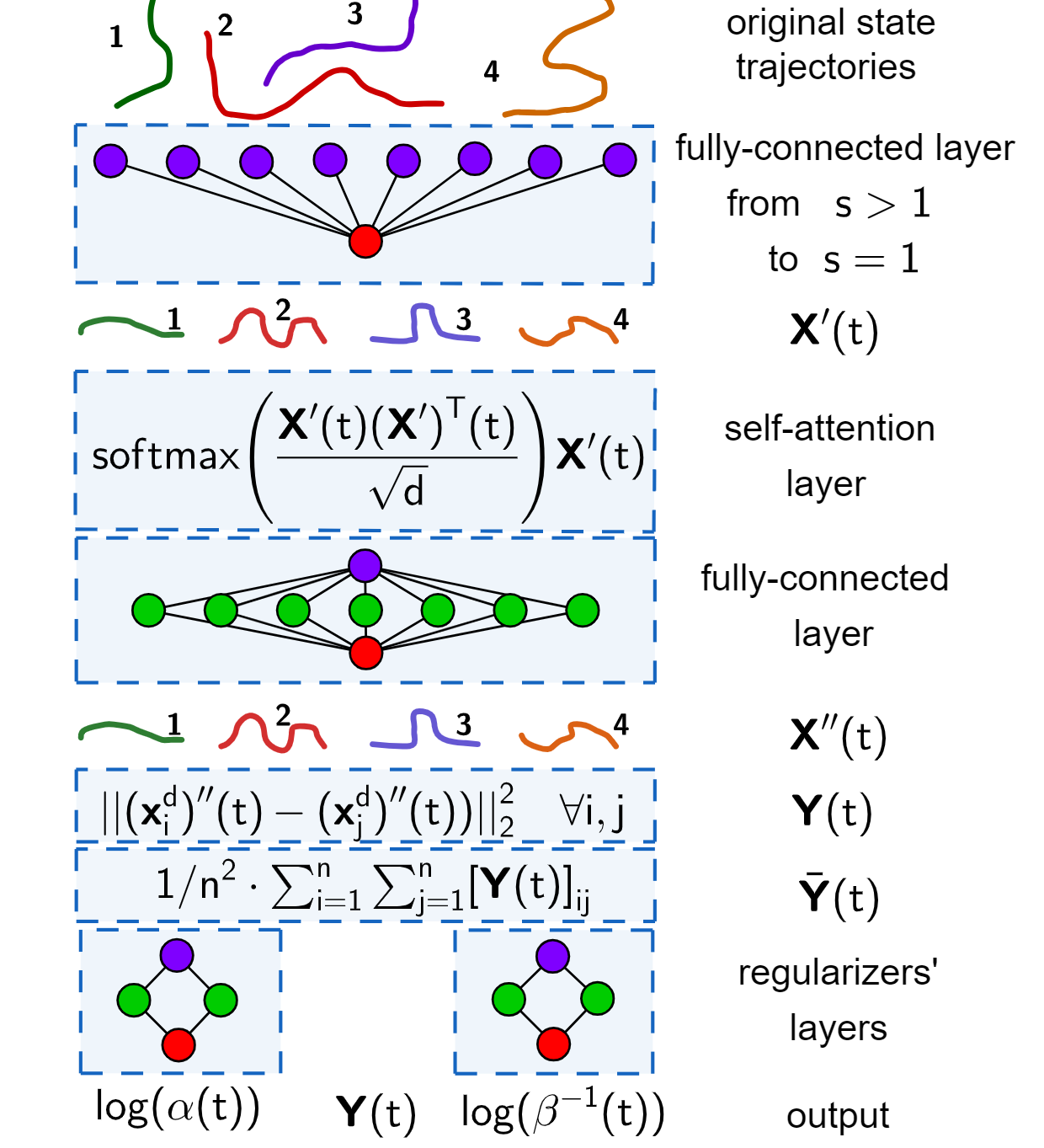}
    \caption{The state trajectory encoder consists of four blocks: (i) a fully connected layer that encodes each individual state to a feature state of dimension $s=1$; (ii) a self-attention layer that finds the relationships among features; (iii) a fully connected series of layers that find $\mathbf{Y}(t)$; and (iv) two separate fully connected layers to find the regularization parameters $\alpha(t)$ and $\beta(t)$ for the subsequent graph identification optimization.}
    \label{fig:encoder}
\end{figure}

The encoder architecture is illustrated  in Fig.~\ref{fig:encoder}. First, a fully connected network projects a node state to a feature state of dimension $s=1$. The input is $\mathbf{x}^d_i(t)$, so the network is applied to every node state and trajectory. The result is a matrix $\mathbf{X}^{\prime}(t)$ that is $n \times d$ as required by the optimization problem in Eq.~\eqref{eq:opt2}. After that, a self-attention layer processes $\mathbf{X}^{\prime}(t)$ according to the relationships found in the attention map $\mathbf{X}^{\prime}(t)(\mathbf{X}^{\prime})^{T}(t)/\sqrt{d}$, using the operation $\mathsf{softmax}(\mathbf{X}^{\prime}(t)(\mathbf{X}^{\prime})^{T}(t)/\sqrt{d})\mathbf{X}^{\prime}(t)$ with $\mathbf{X}^{\prime}(t)$ the query, key and value matrices. Finally, another fully connected network processes every $(\mathbf{x}_i^d)^{\prime}(t)$ to obtain a new matrix $\mathbf{X}^{\prime \prime}(t)$ that is $n \times d$, composed by the individual trajectories $(\mathbf{x}^d_i)^{\prime\prime}(t)$. These feature trajectories are then used to compute the distance matrix $\mathbf{Y}(t)$. \blue{The encoder has an additional module that uses the mean of $\mathbf{Y}(t)$, $\bar{\mathbf{Y}}(t) = \frac{1}{n^2} \sum_{i=1}^{n}\sum_{j=1}^{n}[\mathbf{Y}(t)]_{ij}$, to
compute the parameters $\alpha(t)$ and $\beta(t)$ of Algorithm~\ref{al:smooth}. The logarithm is considered because sparsity is determined by the difference in the order of magnitude between $\alpha(t)$ and $\beta(t)$.}


\blue{The encoder network, thanks to the self-attention structure, handles graphs of different configurations in the number of nodes and the intensity of interactions. Furthermore, it not only projects the state trajectories to a convenient feature space, but also provides the parameters for Algorithm~\ref{al:smooth}. Thus, the encoder can be trained to adjust the connectivity depending on the state trajectories.} For instance, in a multi-robot flocking task like the one used for evaluation in Section~\ref{sec:results}, the robots can depart from a spread initial condition and gather in a more compact formation.
\blue{We remark that the design of the self-attention encoder is not limited to the proposed architecture. Depending on the complexity of the task, the encoder can be increased in depth and number of parameters to ensure that the dimensionality reduction captures all the behaviors of the multi-robot team.}

\subsection{Learning to identify graphs}\label{subsec:proposal}

\begin{figure*}
    \centering
    \includegraphics[width=0.9\textwidth]{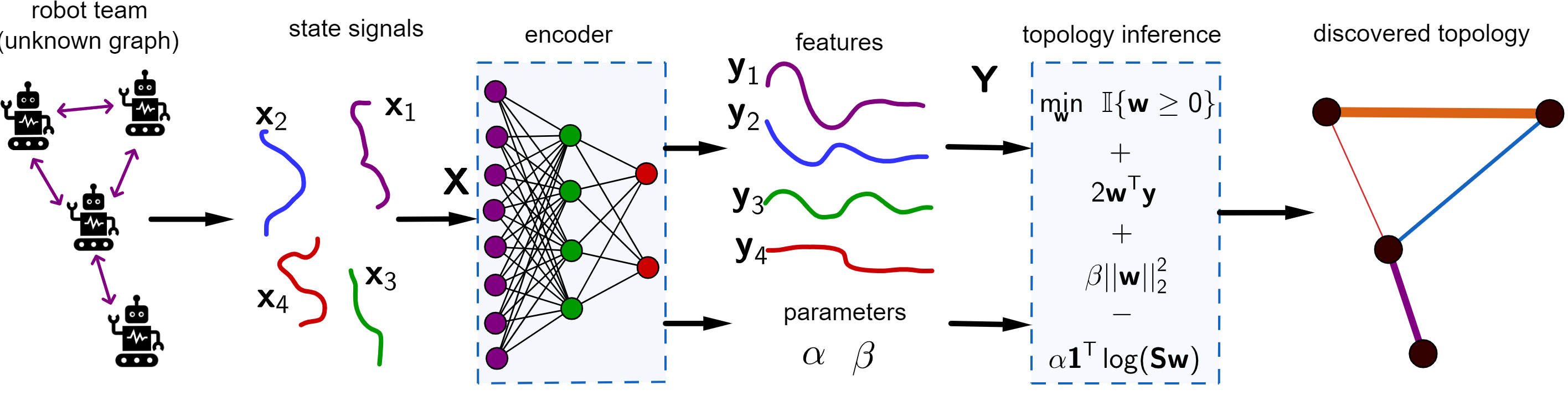}
    \caption{A time-varying graph with unknown connectivity and node dynamics generates a dataset of trajectories (left). A self-attention encoder generates node trajectories in a feature space and computes the regularization parameters. These outputs are the input for an adjacency matrix optimization problem (middle). A fast strongly convex optimization algorithm identifies the weighted adjacency matrix that best describes the observed node trajectories (right).}
    \label{fig:architecture}
\end{figure*}

The combination of the fast convex optimization algorithm described in Section~\ref{subsec:accelerated} and the encoder detailed in Section~\ref{subsec:SAA} leads to a learning architecture for graph topology identification from node state trajectories, presented in Fig.~\ref{fig:architecture}. \blue{The self-attention encoder receives the state trajectories of the multi-robot team. The outputs of the encoder are the one-dimensional feature trajectories of the multi-robot team and the $\alpha$ and $\beta$ regularizers of Algorithm~\ref{al:smooth}. Algorithm~\ref{al:smooth} is then executed using the feature trajectories as input, providing the discovered topology.} To train the model, we use the following loss function:
\begin{equation}\label{eq:graphl_loss_function}
    \mathcal{L}(t) = |\mathbf{w}_k(t) - \hat{\mathbf{w}}| \mathbf{1}_{n(n-1)/2} + |\mathbf{w}^*_k(t) - \hat{\mathbf{w}}^*| \mathbf{1}_{n(n-1)/2}.
\end{equation}
In the loss function, $\hat{\mathbf{w}}$ refers to the vectorized form of the ground-truth weighted adjacency matrix $\bfW$. Moreover, $(\bullet)^*$ denotes the adjoint of a graph. More precisely, $\mathbf{W}^*$ is the weighted adjacency matrix such that $[\mathbf{W}^*]_{ij} = 0$ if $[\mathbf{W}]_{ij} \neq 0$, $[\mathbf{W}^*]_{ij} > 0$ if $[\mathbf{W}]_{ij} = 0$, and $\mathbf{W}^* \mathbf{1}_n = \mathbf{W} \mathbf{1}_n$. We compute each $[\mathbf{W}^*]_{ij} > 0$ as $[\mathbf{W} \mathbf{1}_n]_i/n_i^*$, where $n_i^*$ is the number of non-zero elements of the $i-$th row of $\mathbf{W}^*$. Thus, $\mathbf{w}^*_k $ and $ \hat{\mathbf{w}}^*$ refer to the adjoints of the identified and ground-truth weighted adjacency matrices. The use of the difference of adjoint graphs is to avoid degenerate solutions. For instance, if the ground-truth graph is very sparse, the training might tend to overfit to a graph with no edges unless the adjoint difference is part of the loss.
We consider that each iteration of Algorithm~\ref{al:smooth} is a training step.

Finally, one consideration is in order. Algorithm~\ref{al:smooth} and optimizations~\eqref{eq:opt1}-\eqref{eq:opt2} all provide the optimal graph topology in the state smoothness sense. However, there is one reason for using Algorithm~\ref{al:smooth} instead of~\eqref{eq:opt1} and~\eqref{eq:opt2} beyond the fast convergence. The training of the proposed neural network requires to backpropagate the gradients of the loss function with respect to the output through the complete neural network. In our case, the gradients are propagated backward through the steps of Algorithm~\ref{al:smooth} and the self-attention encoder. Our proposed approach benefits from the fact that these steps are analytical equations rather than an optimization problem as~\eqref{eq:opt1} and~\eqref{eq:opt2}. They are solved by gradient based methods, so, to backpropagate through them, we would need to the gradients of the optimization gradients, which are computationally intractable and numerically unstable. Besides,~\eqref{eq:opt1} has constraints and~\eqref{eq:opt2} as a non-differentiable indicator function, leading to additional difficulties for the training of the neural network.


\section{Evaluation}\label{sec:results}
We conduct two types of experiments\footnote{\scriptsize \url{https://eduardosebastianrodriguez.github.io/LIGMRS/}}. The first experiment (Section~\ref{subsec:static}) considers a multi-robot formation task to verify the ability to extrapolate to other formations and number of robots. The second experiment (Section~\ref{subsec:dynamic}) considers a flocking task to study how the learned neural network extrapolates to other flocking initializations, graph-densities and emergent topologies.

\subsection{Graph identification in multi-robot formation tasks}\label{subsec:static}

Following the setup proposed in many graph identification works \cite{Dong_TSP_2016_LearningSmoothLaplacians, Egilmez_JSTSP_2017_LearningLaplacianGivenAdjency, Giannakis_ProceedingsIEEE_2018_SurveyTopologyIdentificationNonlinearitiesDynamics, Kalofolias_ICASSP_2017_LearningTimeVaryingGraphs, Mei_TSP_2016_SignalProcessingGraphsUnstructuredData,Pavez_TSP_2018_LearningGraphsWithMonotoneTopologyProperties, Saboksayr_EUSIPCO_2021_OnlineGraphLearningSmoothnessPriors,Saboksayr_SPL_2021_AcceleratedGraphLearning, Segarra_TSIPN_2017_NetworkTopologyInferenceSpectralTemplates, Segarra_TSP_2016_BlindIdentificationGraph}, we consider a multi-robot formation problem where the position of the robots is given by an Erd\H{o}s-R\'enyi random graph~\cite{Erdos_PMIHAS_1960_ErdosGraph}. For the training of the neural network, a single graph is instantiated with edge probability $p=0.2$ and a number of robots $n = 50$, \blue{to study the performance of the proposed approach in a sparse and large multi-robot network}. We note that the edge density of the graph $\rho(\mathcal{G}) = p$. Different from other works, instead of considering scalars, the robot state is determined by their 2D position, i.e., $s=2$. The states are generated using the following Gaussian random process:
\begin{equation}
    \kern -0.2cm \mathbf{x}_x(t) \sim \mathcal{N}(\mathbf{0}, \hat{\mathbf{L}}^{\dagger} + \sigma \mathbf{I}_n) \text{ and } \mathbf{x}_y(t) \sim \mathcal{N}(\mathbf{0}, \hat{\mathbf{L}}^{\dagger} + \sigma \mathbf{I}_n),
\end{equation}
where $\hat{\mathbf{L}}^{\dagger}$ is the pseudoinverse of the ground-truth Laplacian, $\sigma = 0.1$ is the noise level \blue{that simulates potential uncertainty in the position of the robots}, and $\mathbf{x}_x(t)$ and $\mathbf{x}_y(t)$ refer to the $x$-position and $y$-position of the robots at instant $t$. The covariance is chosen in this way to force that the position of the robots at each instant is correlated to their neighbors, following the state-of-the-art~\cite{Dong_TSP_2016_LearningSmoothLaplacians, Saboksayr_SPL_2021_AcceleratedGraphLearning} (see \cite{Dong_TSP_2016_LearningSmoothLaplacians} for further details). We generate $d=10000$ samples for training. The other training hyperparameters are detailed in Appendix~\ref{sec:appendix}. To assess the performance of our approach, we randomly generate $18$ test sets composed by $20$ Erd\H{o}s-R\'enyi random graphs each, generated from the combination of the following hyperparameters: $n=\{10, 20, 30, 40, 50, 60\}$ and $p=\{0.1, 0.2, 0.4\}$. All of them are formed by $d=10000$ samples. We set $K=2000$ to ensure convergence. To assess performance, we use the Mean Absolute Error (MAE) between the ground-truth and the identified weights $\frac{1}{n^2}\sum_i^n\sum_j^n|[\hat{\mathbf{W}}]_{ij}-[\mathbf{W}]_{ij}|$. 
We compare our proposal with the state-of-the-art algorithm in~\cite{Saboksayr_SPL_2021_AcceleratedGraphLearning}, which is the closest to our problem assumptions, tuning the parameters according to the procedures detailed in~\cite{Kalofolias_AIS_2016_LearnGraphSmoothSignals, Kalofolias_ICASSP_2017_LearningTimeVaryingGraphs,Saboksayr_EUSIPCO_2021_OnlineGraphLearningSmoothnessPriors}. Since this algorithm only allows scalar trajectories, we apply their algorithm once per state dimension, computing the average graph.
\begin{figure}
    \centering
    \includegraphics[width=0.9\columnwidth]{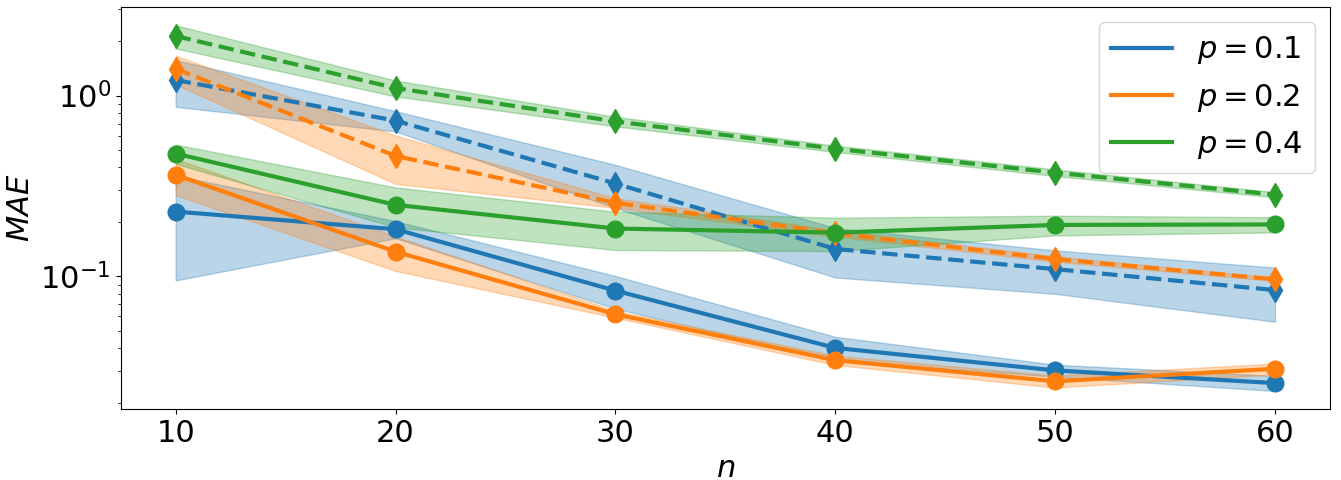}
    \caption{MAE as a function of the number of robots and edge probability. Each configuration is run $20$ times, computing the mean and standard deviation. Diamond dashed lines are the state-of-the-art \cite{Saboksayr_SPL_2021_AcceleratedGraphLearning} ($\alpha=0.2$ and $\beta=0.0001$), whereas the circle solid lines are ours.}
    \label{fig:analysis_n}
\end{figure}

Fig.~\ref{fig:analysis_n} shows the MAE for the different number of robots and edge probabilities. Our proposed approach surpasses the state-of-the-art in one order of magnitude for all the configurations. It is seen how the learned neural network generalizes to different number of robots, achieving a similar performance in terms of MAE.  \blue{Therefore, the training has been able to learn a good encoding of the state trajectories and identifies a good $\alpha$ and $\beta$.} We emphasize that the training is conducted with just one graph. The learned neural network does not generalize to $\rho(\mathcal{G}) =p=0.4$ since only a single graph with $\rho(\mathcal{G}) = p = 0.2$ has been used for training.

\begin{figure}
    \centering
    \begin{tabular}{cccc}
{\small $\hat{\mathbf{W}}$}
&
{\small ${\mathbf{W}}$}
&
{\small $\hat{\mathbf{W}} - {\mathbf{W}}$}
&
{\small threshold}
    \\
\includegraphics[width=0.25\columnwidth, height=0.3\columnwidth] {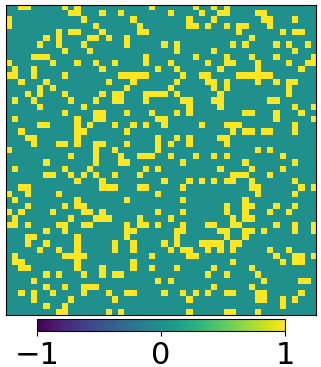}
         &
\includegraphics[width=0.25\columnwidth, height=0.3\columnwidth] {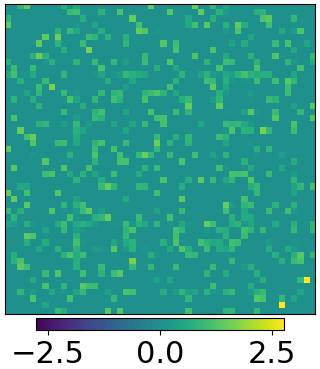}
         &
\includegraphics[width=0.25\columnwidth, height=0.3\columnwidth] {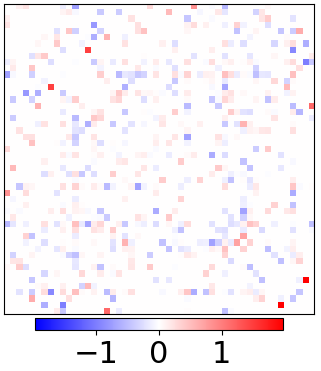}
         &
\includegraphics[width=0.25\columnwidth, height=0.3\columnwidth] {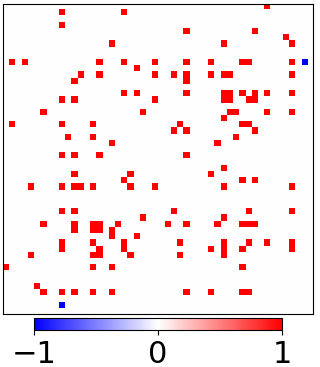}
    \end{tabular}
    \caption{After the training, the learned neural network correctly identifies the ground-truth graph. The intensity of the colors in the discovered graphs is lower because there are a few outlier weights with a greater value compared to the ground-truth. The right pannel shows the difference between identified and ground-truth weighted adjacency matrices, but with a threshold $|[\hat{\mathbf{W}}]_{ij}-[\mathbf{W}]_{ij}|<10^{-5}$. }
    \label{fig:learned_graph}
\end{figure}

\begin{figure}
    \centering
    \begin{tabular}{cccc}\vspace{-0.1cm}
{\small $\hat{\mathbf{W}}$}
&
{\small ${\mathbf{W}}$}
&
{\small $\hat{\mathbf{W}} - {\mathbf{W}}$}
&
{\small threshold}
    \\\vspace{-0.1cm}
\includegraphics[width=0.25\columnwidth, height=0.3\columnwidth] {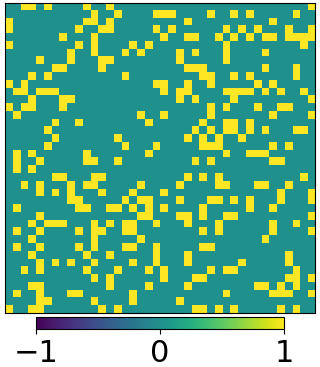}
         &
\includegraphics[width=0.25\columnwidth, height=0.3\columnwidth] {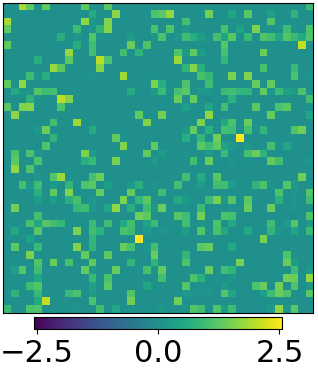}
         &
\includegraphics[width=0.25\columnwidth, height=0.3\columnwidth] {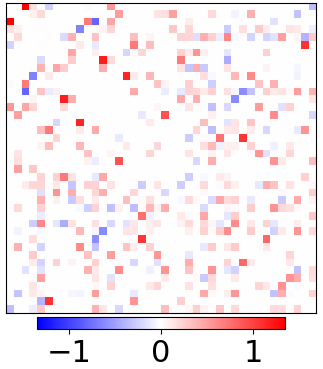}
         &
\includegraphics[width=0.25\columnwidth, height=0.3\columnwidth] {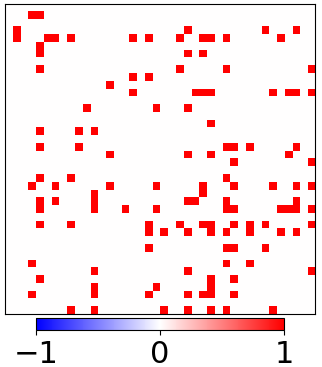}
    \\\vspace{-0.1cm}
\includegraphics[width=0.25\columnwidth, height=0.3\columnwidth] {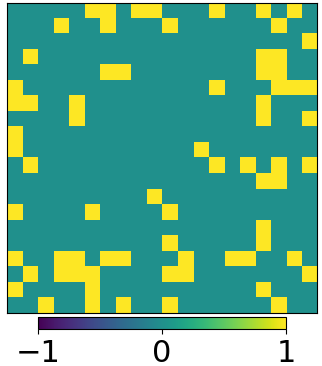}
         &
\includegraphics[width=0.25\columnwidth, height=0.3\columnwidth] {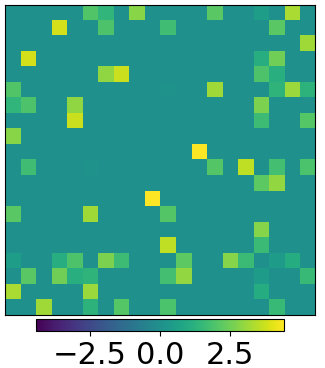}
         &
\includegraphics[width=0.25\columnwidth, height=0.3\columnwidth] {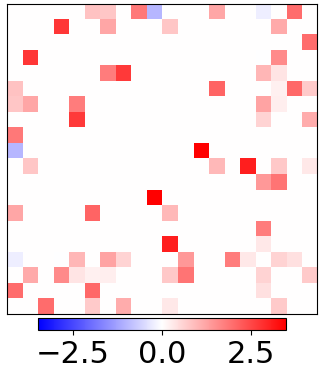}
         &
\includegraphics[width=0.25\columnwidth, height=0.3\columnwidth] {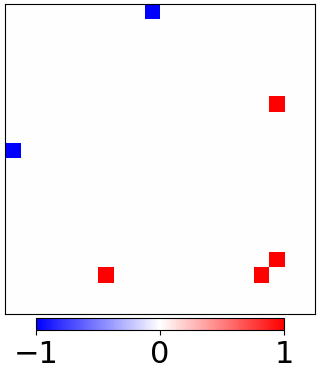}
    \\ 
\includegraphics[width=0.25\columnwidth, height=0.3\columnwidth] {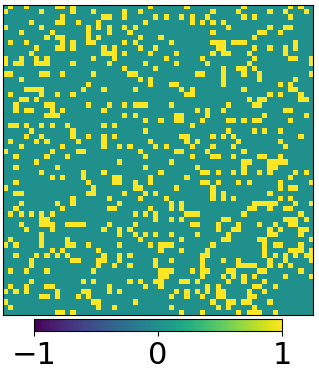}
         &
\includegraphics[width=0.25\columnwidth, height=0.3\columnwidth] {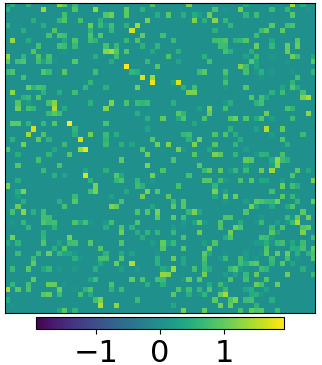}
         &
\includegraphics[width=0.25\columnwidth, height=0.3\columnwidth] {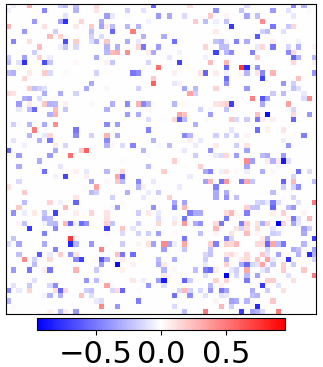}
         &
\includegraphics[width=0.25\columnwidth, height=0.3\columnwidth] {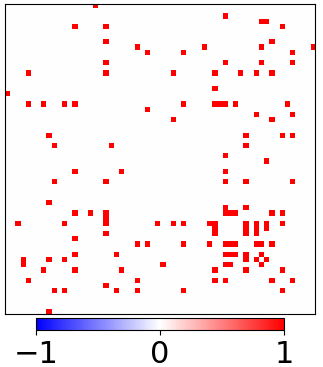}
    \end{tabular}
    \caption{Qualitative results obtained by our approach for different number of robots. Our approach is accurate in the identification of the weights irrespective of the number of robots.}
    \label{fig:scalability}
\end{figure}

\begin{figure}
    \centering
    \begin{tabular}{cccc}
{\small $\hat{\mathbf{W}}$}
&
{\small ${\mathbf{W}}$}
&
{\small $\hat{\mathbf{W}} - {\mathbf{W}}$}
&
{\small threshold}
    \\\vspace{-0.1cm}
\includegraphics[width=0.25\columnwidth, height=0.3\columnwidth] {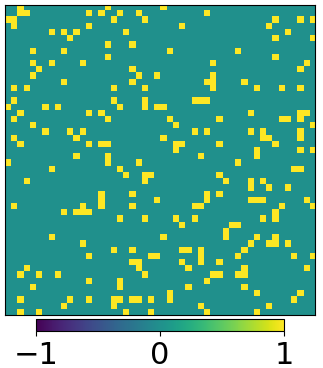}
         &
\includegraphics[width=0.25\columnwidth, height=0.3\columnwidth]{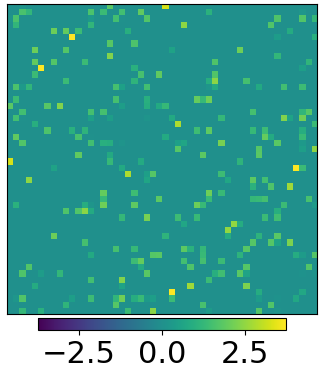} 
         &
\includegraphics[width=0.25\columnwidth, height=0.3\columnwidth]{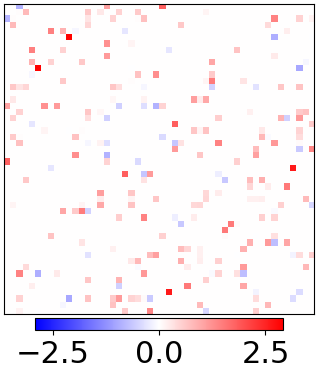} 
         &
\includegraphics[width=0.25\columnwidth, height=0.3\columnwidth]{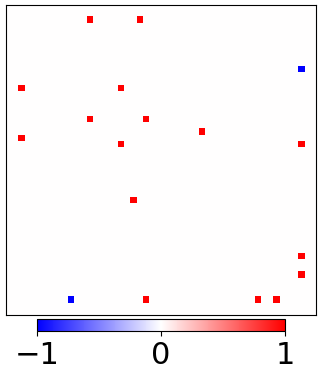} 
    \\\vspace{-0.1cm}
\includegraphics[width=0.25\columnwidth, height=0.3\columnwidth] {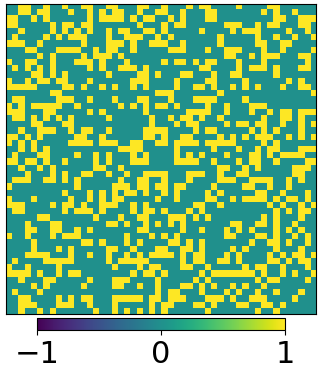}
         &
\includegraphics[width=0.25\columnwidth, height=0.3\columnwidth] {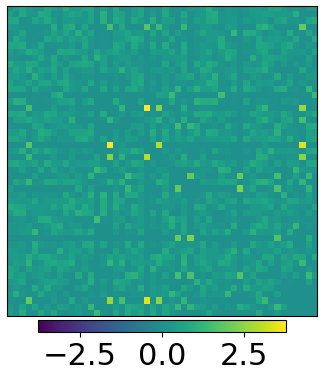}
         &
\includegraphics[width=0.25\columnwidth, height=0.3\columnwidth] {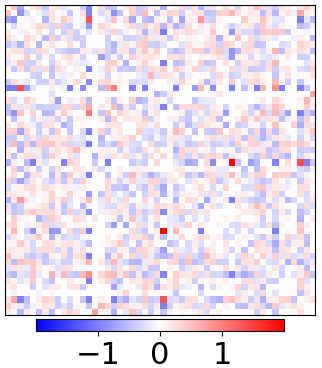}
         &
\includegraphics[width=0.25\columnwidth, height=0.3\columnwidth]{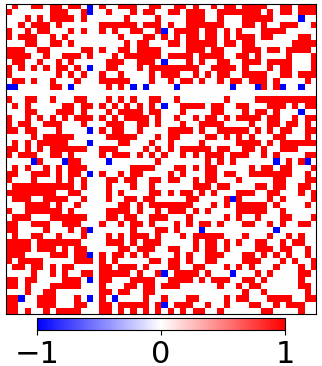} 
    \end{tabular}
    \caption{Our approach moderately generalizes to other edge densities during evaluation because the training set only consists of a single graph with $p=0.2$. Top row shows a case with $p=0.1$, and the bottom row shows a case with $p=0.4$.}
    \label{fig:sparsity}
\end{figure}

Figs. \ref{fig:learned_graph}, \ref{fig:scalability} and \ref{fig:sparsity} show some qualitative results. The weights of the existing edges are accurately identified in graphs with a similar number of robots than in training (Fig. \ref{fig:learned_graph}), different numbers of robots (Fig. \ref{fig:scalability}) and edge densities (Fig. \ref{fig:sparsity}). The intensity of the colors in the discovered graphs is lower than in the ground-truth graphs because there are a few outlier weights with a greater value compared to the ground-truth, so the color scales of $\mathbf{W}$ and $\hat{\mathbf{W}}$ are different. Since the $\hat{\mathbf{W}}$ from the Erd\H{o}s-R\'enyi graphs are such that $[\hat{\mathbf{W}}]_{ij} = \{0,1\}$ and to verify that our approach is successful at capturing the existence or absence of edges, the right panels of Figs. \ref{fig:learned_graph}, \ref{fig:scalability} and \ref{fig:sparsity} show the difference between identified and ground-truth weighted adjacency matrices, but thresholding to zero all the elements such that $|[\hat{\mathbf{W}}]_{ij}-[\mathbf{W}]_{ij}|<10^{-5}$. \blue{Note that this threshold is to remove edges whose order of magnitude is far from the weights of an existing edge}. With a correct threshold, the existence or absence of edges is correctly identified for all configurations, except for the case of $\rho(\mathcal{G}) = p = 0.4$, where the number of edges is underestimated because only a single graph with $\rho(\mathcal{G}) = p = 0.2$ has been used for the training of the neural network. 


\subsection{Graph identification in multi-robot flocking tasks}\label{subsec:dynamic}

The next experiment evaluates our approach when it is trained with a variety of graph edge densities. We study our proposed approach in a multi-robot flocking problem, where the state is defined by the 2D position the robots, so $s=2$. Robot trajectories are generated using the controller proposed in~\cite{olfati2006flocking}, parameterized as detailed in Appendix~\ref{sec:appendix}. The training set is generated with a sparsity pattern determined by the desired inter robot distance $\rho = 0.7$m and communication radius $r_{comm} = 1.2$m, simulated through $6$s with a sample time of $0.04$s, and where the robots are uniformly spawned in a square of $5\times 5$m. The trajectories are then split in sub-trajectories of $d=10$, each of them associated to a graph resulting from the average of the $d$ samples. This leads to a training set of graphs with edge densities from $5\%$ to $95\%$. The test cases are generated with the following configurations, one trajectory each: (1) $\rho = 0.7$m and $r_{comm} = 1.2$m; (2) $\rho = 1.0$m and $r_{comm} = 1.2$m; and (3) $\rho = 0.7$m, $r_{comm} = 1.2$m and a compact initialization in a square $2\times2$m. The desired position of the flock is always $x^* = y^*=0.0$m. The training of the neural network takes $150000$ steps. Every $500$ steps, one graph with its associated sub-trajectories is uniformly randomly picked, initializing the optimization module each time. 

\begin{figure}
    \centering
    \includegraphics[width=0.9\columnwidth]{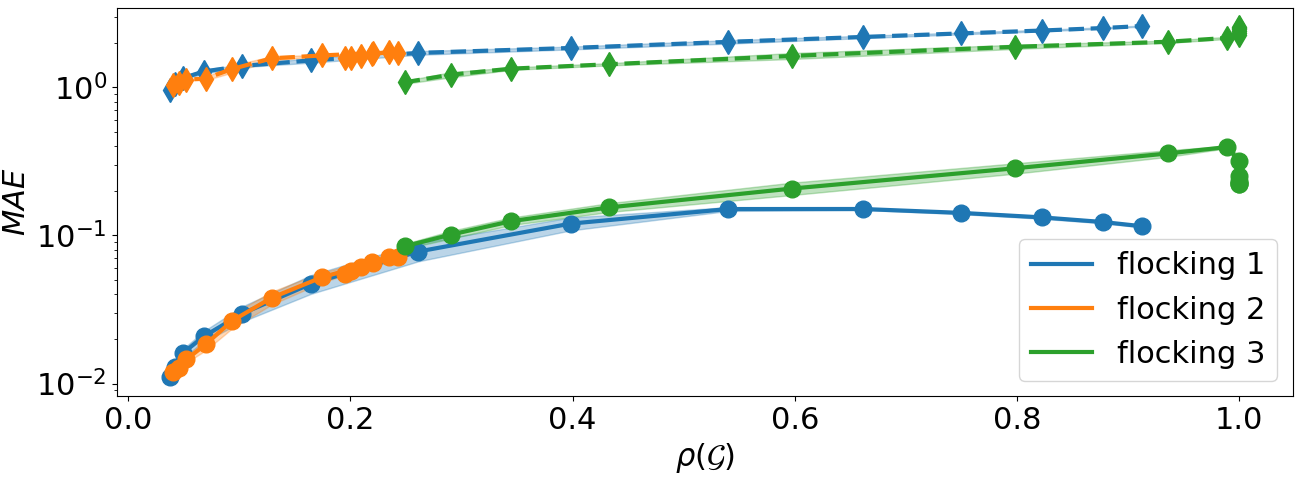}
    \caption{MAE as a function of the edge density and flocking configurations. Each configuration is run $10$ times, computing the mean and standard deviation. Diamond dashed lines are the state-of-the-art \cite{Saboksayr_SPL_2021_AcceleratedGraphLearning} ($\alpha=0.1$ and $\beta=10^{-5}$), whereas the circle solid lines are ours.}
    \label{fig:analysis_flocking}
\end{figure}


\begin{figure}
\setlength{\tabcolsep}{7pt}
    \centering
    \begin{tabular}{ccc}\vspace{-0.1cm}
{\small $\hat{\mathbf{W}}$}
&
{\small ${\mathbf{W}}$}
&
{\small $\hat{\mathbf{W}} - {\mathbf{W}}$}
    \\\vspace{-0.1cm}
\includegraphics[width=0.28\columnwidth, height=0.3\columnwidth] {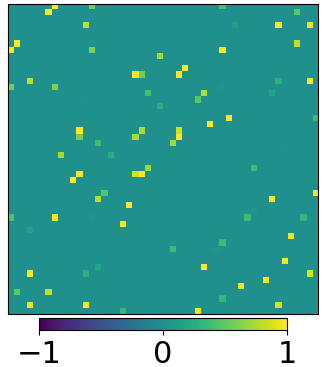}
         &
\includegraphics[width=0.28\columnwidth, height=0.3\columnwidth] {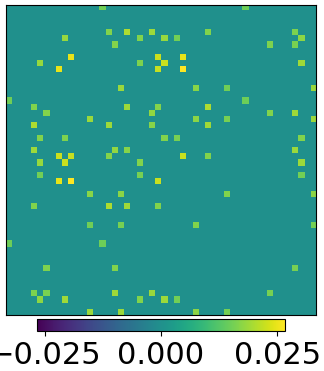}
         &
\includegraphics[width=0.28\columnwidth, height=0.3\columnwidth] {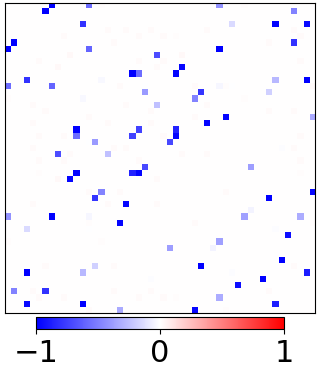}
    \\\vspace{-0.1cm}
\includegraphics[width=0.28\columnwidth, height=0.3\columnwidth] {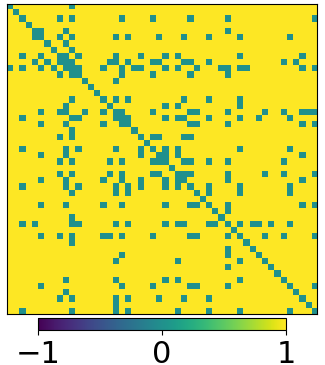}
         &
\includegraphics[width=0.28\columnwidth, height=0.3\columnwidth] {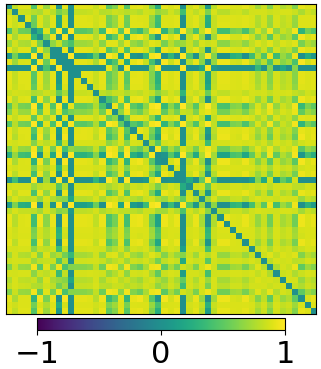}
         &
\includegraphics[width=0.28\columnwidth, height=0.3\columnwidth] {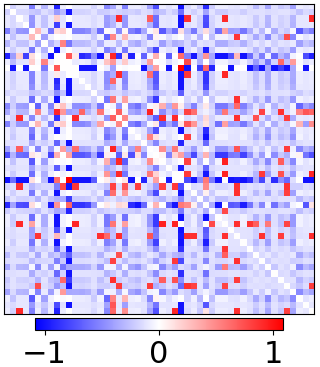}
\\
\includegraphics[width=0.28\columnwidth, height=0.3\columnwidth] {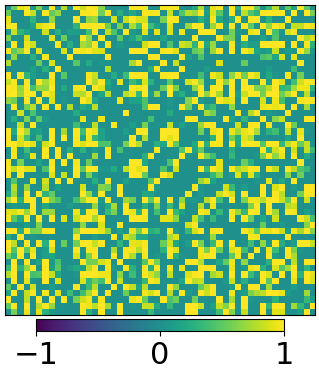}
         &
\includegraphics[width=0.28\columnwidth, height=0.3\columnwidth] {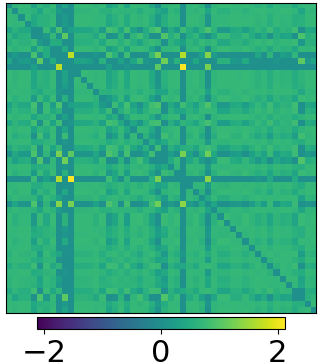}
         &
\includegraphics[width=0.28\columnwidth, height=0.3\columnwidth] {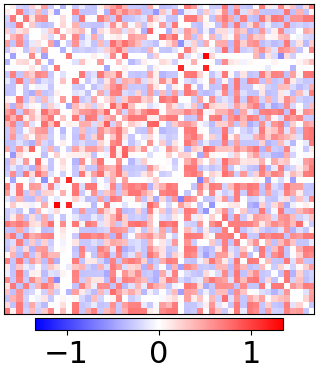}
    \end{tabular}
    \caption{Qualitative results of our approach for different flocking configurations and edge densities, extracted from the three flocking examples. The relative value among edges of the discovered weights is accurately identified, according to the color scales. }
    \label{fig:sparsity_flocking}
\end{figure}

The results in Fig. \ref{fig:analysis_flocking} show how the MAE evolves as a function of the edge density and flocking configuration. For all the cases, our approach outperforms the state-of-the-art by more than one order of magnitude. Compared to the formation problem, the difference now is greater because the dynamics of the robots under the flocking controller are nonlinear. \blue{Moreover, in the state-of-the-art algorithms, $\alpha$ and $\beta$ are static, so the output is not able to adjust to changes in edge density. In contrast, our approach adjusts the optimization regularizers and provides the latent feature trajectories that best fit the flocking task, concluding that our approach has learned to adapt the parameters of the optimization method to the observed edge probability/density}. Compared to the formation task, the learned neural network achieves a good performance for the different edge densities because the training set is diverse in this aspect. The MAE in Fig.~\ref{fig:analysis_flocking} is specially good in very sparse networks ($\rho(\mathcal{G})<0.25$). As observed in Fig.~\ref{fig:sparsity_flocking}, the relative value among the discovered weights is correctly identified, whereas the value of these weights with respect to the ground-truth $\hat{\mathbf{W}}$ vary depending on the edge density of the graph. Looking at Fig. \ref{fig:sparsity_flocking}, for sparse, medium and dense graphs the identified weights tend to be lower, similar and greater than in the ground-truth graph. This also explains why the MAE in Fig.~\ref{fig:analysis_flocking} grows with the edge density. Besides, the neural network predicts a greater number of cliques, as in the middle row of Fig. \ref{fig:sparsity_flocking}. These behaviors require further investigation. \blue{As future work, We also plan to conduct ablation studies to analyze the impact of the dataset edge density variety in the generalization capabilities of our approach}.    


\section{Conclusion}\label{sec:conclusions}
This work proposed a novel approach for graph topology identification combining a self-attention encoder with a fast analytical convex optimization algorithm. Our method provides a neural network model that learns to identify general graph topologies using only state trajectories of the nodes. Our approach is accurate and flexible, surpassing the state-of-the-art in multi-robot graph identification with different node configurations, node number, and edge densities. The applications to multi-robot systems problems are diverse. For instance, it can be used to identify the topology in multi-agent demonstrations to constrain the learning of multi-robot control policies or to detect failures in the communications among warehouse robots. \blue{In addition, by considering agents instead of robots, the proposed approach can be directly applied to general networked systems like brain imaging, genetics or social interactions.}



\appendices

\section{Hyperparameters}\label{sec:appendix}
The training of the neural networks evaluated in Section~\ref{sec:results} is parameterized by a learning rate $\mu = 0.001$ and uses Adam \cite{kingma2014adam}. Algorithm~\ref{al:smooth} is parameterized by $\epsilon = 10^{-5}$. The encoder of the neural network in Section~\ref{subsec:static} is given by a first fully-connected network of layer dimensions $[2, 5, 10, 5, 1]$ with \textsf{tanh} activation functions and a second fully connected network of layer dimensions $[1, 2, 1]$ with \textsf{tanh} activation functions except the last one, which is linear. The layers for $\alpha$ and $\beta$ are of dimensions $[1, 2, 1]$ with \textsf{tanh} activation functions except the last layer, which is a sigmoid. The encoder of the neural network in Section~\ref{subsec:dynamic} is given by a first fully-connected network of layer dimensions $[2, 4, 4, 1]$ with \textsf{tanh} activation functions. The layers for $\alpha$ and $\beta$ are of dimensions $[1, 2, 1]$ with \textsf{tanh} activation functions except the last layer, and the output is multiplied by a scalar $b=3$. The flocking follows the dynamics proposed in~\cite{olfati2006flocking}, with parameters: $\epsilon = 0.1$, $a=5.0$, $b=5.0$, $h = 0.2$, $c_1 = 0.4$, $c_2 = 0.8$.

\balance
\bibliographystyle{IEEEtran}
\bibliography{IEEEabrv,IEEEexample.bib}

\end{document}